# Cumulenic *sp*-Carbon Atomic Wires wrapped with Polymer for Supercapacitor Application

Subrata Ghosh[1*], Massimiliano Righi[1], Simone Melesi[1], Yu Qiu,[2] Rik R. Tykwinski[2], Carlo S. Casari[*1]

[1] *Micro and Nanostructured Materials Laboratory — NanoLab, Department of Energy, Politecnico di Milano, via Ponzio 34/3, Milano, 20133, Italy*

[2] *Department of Chemistry, University of Alberta, Edmonton, Alberta, Canada T6G 2G2*

ABSTRACT:

Carbon atomic wires, a linear atomic chain of *sp*-carbon, is theoretically predicted to have around five times higher surface area than graphene, notable charge mobilities, as well as excellent optical and thermal properties. Despite these impressive properties, the properties of *sp*-carbon as an electrochemical energy-storage electrode have not been reported so far. Herein, we prepare solution-processed thin films of tetraphenyl[3]cumulenic *sp*-carbon atomic wires embedded in a polymer matrix, in which sp-carbon atomic wires feature three cumulated carbon-carbon double bonds terminated at each end by two phenyl groups. Raman and UV-visible spectroscopy are used to confirm the presence and possible degradation of *sp*-carbons inside the polymeric matrix. Finally, we investigate the supercapacitor performance of cumulenic *sp*-carbon atomic wires embedded polymer in three aqueous mediums, namely 1M $Na_2SO_4$ (neutral), 1M $H_2SO_4$ (acidic), and 6M KOH (basic). The results suggest 6M KOH is the best electrolyte to obtain high charge-storage performance of device with areal capacitance of 2.4 mF/cm$^2$ at 20 mV/s, 85% cycle stability after 10000 charge-discharge cycles, and excellent frequency response.

Corresponding author email: subrata.ghosh@polimi.it (S.G.) and carlo.casari@polimi.it (C.S.C.)

ORCID ID: 0000-0002-5189-7853 (S.G.), 0009-0009-8268-6926 (S.M.), 0000-0002-7645-4784 (R.R.T.), 0000-0001-9144-6822 (C.S.C.)

INTRODUCTION

The electrochemical capacitor or supercapacitor is anticipated as a promising electrochemical energy device to store the charge from renewable energy sources as well as high-power applications that require rapid on–off response such as hybrid electric vehicles, forklifts, load cranes, grid stabilization systems, etc.[1] Since the charge-storage mechanism is completely physical, unlike the chemical storage mechanism in batteries, a supercapacitor is well-known for the excellent power density, cycle stability, safety, and capability to work in a broad temperature range (–10 to 100°C). However, low energy density restricts its use only for high-power applications. To improve the energy density of supercapacitors, one can increase the capacitance by tuning the porosity and structural properties of electrode materials, doping or decorating carbon-nanostructures, use electrolytes to increase the voltage of device, and also choosing the best electrode-electrolyte combination to obtain better charge-storage performances.[2][3][4]

There is extensive research on materials as supercapacitor electrodes starting from nanocarbons, 2D materials (e.g., MXene, transition metal dichalcogenides), conducting polymers, and their composites. [2][3] [4][5][6][7] Amongst these, carbon-based materials come to the spotlight due to their high surface area, controllable porosity, excellent structural properties like electrical conductivity and thermal conductivity, and excellent stability. On the other hand, limited density of states, chemical inertness, low packing density, and poor wettability are the limiting factors to obtain higher storage performance. Among the carbon-based materials, although graphene, which is 100% $sp^2$-carbon, has high surface area (2630 m$^2$/g) and excellent properties, the predicted specific capacitance of 550 F/g was not achieved so far. The KOH activation of microwave exfoliated graphene oxide (98% $sp^2$-carbon) is explored as an excellent supercapacitor electrode. With specific surface area of 3100 m$^2$/g, the microwave exfoliated graphene oxide electrode delivered the gravimetric capacitance of 200 F/g at the current density of 0.7 A/g[8] However, the small overpotential of oxygen and hydrogen evolution reaction of $sp^2$-carbons in aqueous electrolyte often restrict the operating voltage window within 1.23V during the charge-discharge process. On the other hand, diamond ($sp^3$-bonded carbon) is promising and can operate in widened potential window both in aqueous and non-aqueous electrolytes with excellent electrochemical stability.[9][10] Unfortunately, however, the specific capacitance of diamond-based structures is quite low. Thus, the attention has turned to hybrid structures of $sp^2$-$sp^3$ carbon as materials energy storage electrode. It has been reported, for example, that the Q-carbon microdots (81.3% $sp^3$- carbon delivers higher charge-storage performance than other carbon structures, including Q-carbon filament (78.3% $sp^3$-carbon) and Q-carbon cluster (73% $sp^3$-carbon).[11]

Apart from $sp^2$-carbon and $sp^3$-carbon, $sp$-carbon offers other forms of carbon that exist for materials applications. It has been predicted that the ideal infinite linear chain of $sp$-carbons, has a high theoretical surface area of more than 13000 m$^2$/g for H$_2$ storage, which is much greater than graphene (2630 m$^2$/g).[12] Carbyne should also possesses excellent thermal conductivity (80 ± 26 KW/m/K at room temperature)[13] and electronic mobility (>10$^5$ cm$^2$/Vs)[14][15]. The linear chain of $sp$-carbons has two limiting structures based on the bonding arrangement in the linear $sp$-carbon chain: (a) polyynes with alternating single and triple carbon-carbon bonds and a semiconducting behavior and (b) cumulenes with contiguous double bonds and metallic behavior.[16] Experimentally, infinite strands of carbyne are not synthesizable due their reactivity. Thus, $sp$-carbon chains have been obtained only in the form of carbon atomic wires (CAWs) with finite lengths and with different endgroups (e.g., H, CN, methyl, aryl, halogen).[17][18][19] The choice of the chain length and the end-group are fundamental since both can affect the stability and the tunability of the optoelectronic properties of these $sp$-carbon systems. The optical energy gap can be modulated by varying the number of *sp*-carbons in the chain (i.e., longer chains present a lower optical gap) and by changing the endgroups.[17][20][21] Bulkier endgroups increase the stability of these systems.[22] Another possibility to protect CAWs is their embedding inside polymeric matrices[23][24][25] avoiding exposure to detrimental environmental agents, reducing their mobility, and diminishing the tendency of crosslinking among adjacent *sp*-carbon chains that leads to more stable $sp^2$ structures.[26][27][28]

Carbon-based materials with an increasing proportion of *sp*-hybridized are attractive to researchers for hydrogen storage,[12] electrochemical energy storage,[29] field effect transistors,[30] and organic electronic applications.[31] It has been reported that the *sp*-carbon content in nanothick amorphous carbon coating has strong influence on its mechanical properties.[32] The nanostructures rich *sp*-carbon have been explored as a supercapacitor electrode in ionic electrolyte and the maximum areal capacitance of 0.32 mF/cm$^2$ at 0.05 V/s is obtained with respect to the reference electrode.[29] In another study, carbon coated Ni foam is reported to deliver the areal capacitance of 53.06 mF/cm$^2$ at 5 mV/s in 1M Na$_2$SO$_4$ with respect to the reference electrode.[33] This latter study,[33] however, presents several concerns: (i) Reporting the specific capacitance below 20 mV/s may not be appropriate, (ii) The shape of the CV traces is not ideal for supercapacitor, and, more importantly, (III) the signal from *sp*-carbon was not identified from Raman spectra. In the report,[30] the Raman peaks at 1143.5 cm$^{-1}$ and 1539.8 cm$^{-1}$ are attributed to the β-carbyne (cumulenic carbyne), and carbon networks (*sp*-, $sp^2$-, and $sp^3$-carbon), respectively. However, the main peaks in Raman spectra for *sp*-carbon chains should be in the range of 1800 and 2300 cm$^{-1}$.[34] In our previous study,[35] we investigated the carbon nanofoam containing *sp*-, $sp^2$-, and $sp^3$-carbon with hydrogen, nitrogen, and oxygen functionalization for supercapacitor applications. However, the exact role of *sp*-carbon of carbon nanofoam on the supercapacitor performance is not clear. Despite the promising features, there are no reports on the *sp*-carbon based structure as an electrochemical energy storage electrode to the best of our knowledge.

In this work, we chose a CAW with three cumulated carbon-carbon double bonds terminated at each end by two phenyl groups ([3]Ph-cumulene). We produced a polymeric thin film embedding [3]Ph-cumulene into poly(methyl methacrylate) (PMMA), and we investigated the charge-storage performance and frequency response of cumulenic *sp*-carbon atomic wires-based symmetric supercapacitor devices in three different aqueous electrolytes (1M Na$_2$SO$_4$, 1M H$_2$SO$_4$ and 6M KOH). We also investigated the structural stability of [3]Ph-cumulene after electrochemical measurements by Raman spectroscopy.

RESULTS & DISCUSSIONS

The electrode material using [3]Ph-cumulene is prepared by embedding the [3]Ph-cumulene within a poly(methyl methacrylate) drop casted thin film on a carbon paper substrate (Scheme 1, see Experimental section for further details). The symmetric device is assembled by two identical electrodes and separator-soaked-electrolyte.

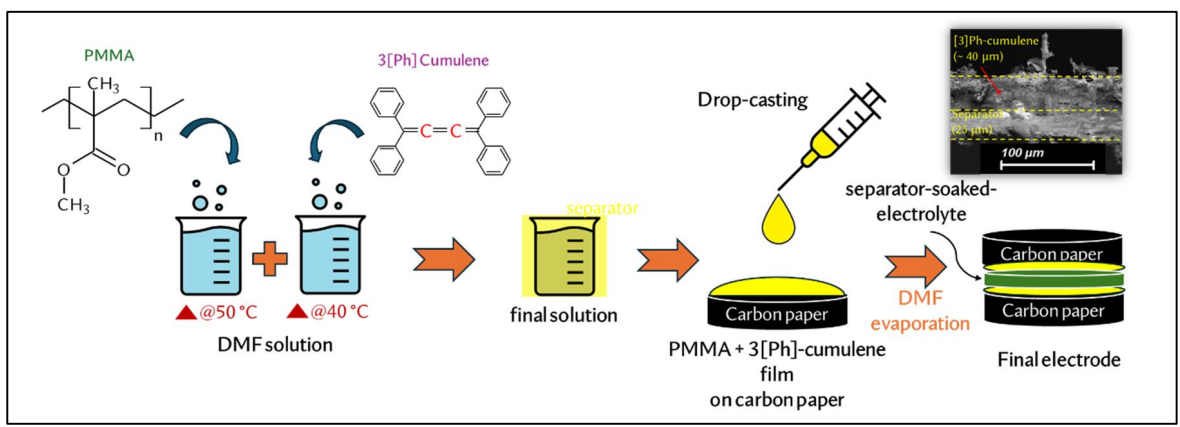

*Scheme 1: Schematic of [3]Ph-cumulene-PMMA electrode fabrication. The right top image is the cross-sectional scanning electron micrographs of thin film electrode on separator.*

To ensure the correct embedding and to verify that the [3]Ph-cumulene retained its properties during processin, Raman spectroscopy has been performed on the PMMA powder, [3]Ph-cumulene powder, and on the [3]Ph-cumulene film. The spectra are shown in Figure 1a. The spectrum of PMMA is characterized mainly by three peaks in the region of 2800–3000 cm$^{-1}$ that are associated with the CH stretching of the polymeric chain. The Raman spectra of [3]Ph-cumulene consist of a strong peak at 2035 cm$^{-1}$ related to the Effective Conjugation Coordinate (ECC) normal mode, a collective vibration of all the *sp*-carbon atoms in the chain.[21] At 1593 cm$^{-1}$ the [3]Ph-cumulene presents another remarkable peak associated with the stretching mode of the phenyl endgroups. In the spectrum of the [3]Ph-cumulene-PMMA film, the cumulenic Raman modes remain prominent and suggest that the cumulene is stable and degradation does not occur during the preparation of the samples. Interestingly, in this spectrum, the PMMA peaks are barely distinguishable. This can be explained both considering the high concentration of cumulene powder in the films and the higher Raman scattering cross-section of cumulenes over the polymer.[36] From the UV-vis absorption spectra, we verified the structure of [3]Ph-cumulene with three absorption bands observed at 415, 317, and 270 nm (Figure 1b).[30][37] Notably, the UV-vis spectra of [3]Ph-cumulene and [3]Ph-cumulene + PMMA show nearly identical profiles suggesting that PMMA does not affect the properties of [3]Ph-cumulene.

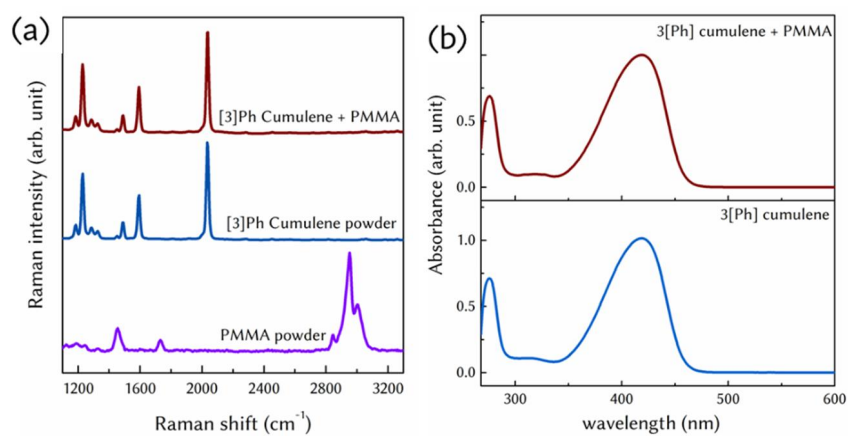

*Figure 1 (a) Raman spectra of PMMA powder (purple), [3]Ph-cumulene powder (blue), and drop casted thin film of [3]Ph-cumulene + PMMA (red). (b) UV-Vis absorbance spectra of the [3]Ph-cumulene (in DMF) and [3]Ph cumulene + PMMA (in DMF); DMF: N,N-dimethylformamide*

The charge-storage performance of [3]Ph-cumulene-PMMA composite supercapacitor devices is investigated in three different aqueous electrolytes, namely $H_2SO_4$ (acidic), $Na_2SO_4$ (neutral), and KOH (basic). Figure 2a shows the representative cyclic voltammograms (CVs) of symmetric device with 1M $Na_2SO_4$ electrolyte at different scan rate in the range of 0.2 mV/s to 1 V/s. The CV analyses of the devices in 1M $H_2SO_4$ and 6M KOH are shown in Figure S1a–b. The CV profile is found to be near-rectangular and mirror symmetric for all scan rates indicating excellent supercapacitor behaviour. The noticeable difference in the CV for different electrolytes is their capability to work at different voltages (Figure 2b). The plot of areal capacitance versus scan rate for the PMMA/6M KOH/PMMA symmetric device is provided in Figure S1c, and the areal capacitance is determined to be much lower than the cumulene-based symmetric device at the given scan rate. The composite/$Na_2SO_4$ system can sustain the voltage of 1 V, the stable voltages of devices for 1M $H_2SO_4$ and 6M KOH is found to be 0.8 V without observable evolution of oxygen and hydrogen. At 0.1 V/s, the energy density of all devices is found to be around 0.11 µWh/cm$^2$. This result suggests that the use of $Na_2SO_4$ electrolyte may not be a great option despite a widened stable voltage is achieved. Thus, it is important to inspect other parameters like rate performance, cycle stability, and charge-storage kinetics of the electrode under different electrolytes. The areal capacitance of the devices in different aqueous electrolytes is estimated and plotted with respect to the scan rate (Figure 2c). The highest areal capacitance of device is obtained in 6M KOH (2.4 mF/cm$^2$) and in $H_2SO_4$ (2.8 mF/cm$^2$), whereas the lowest of is observed in $Na_2SO_4$ (2 mF/cm$^2$), with all measurements at the scan rate of 0.02 V/s. This higher specific capacitance of the composite in KOH and $H_2SO_4$ electrolyte over $Na_2SO_4$ can be attributed to the lower ionic radius and higher ionic mobility of $H^+$-ions for composite/$H_2SO_4$, and better wettability of KOH and higher $K^+$-ion diffusion for composite/KOH.[38] Our composite (areal capacitance of electrode = 4 × specific capacitance of device = 5.2 mF/cm$^2$ at 0.1 V/s) outperforms values reported for vertical graphene nanosheets (0.197 mF/cm$^2$),[39] $Ni_3(HHTP)_2$@woven fabrics (0.205 mF/cm$^2$), $TiO_2$ nanogrids (0.74 mF/cm$^2$),[40] and titanate hydrate nanogrids (0.08 mF/cm$^2$) (Table 1).[40] Specific to the *sp*-carbon containing nanostructures of [3]Ph-cumulene, the areal capacitance of our composite is found to be 2.8 mF/cm$^2$ in 6M KOH at 1 V/s, which is higher than the previously reported *sp*-carbon-rich nanostructured materials (264 µF/cm$^2$) and carbyne-depleted nanostructured carbon (234 µF/cm$^2$) at 1 V/s for in [Emim][NTf2] electrolyte in three-electrode system (both values are derived here from data extracted from the plots in reference [29] using webplotdigitizer software).

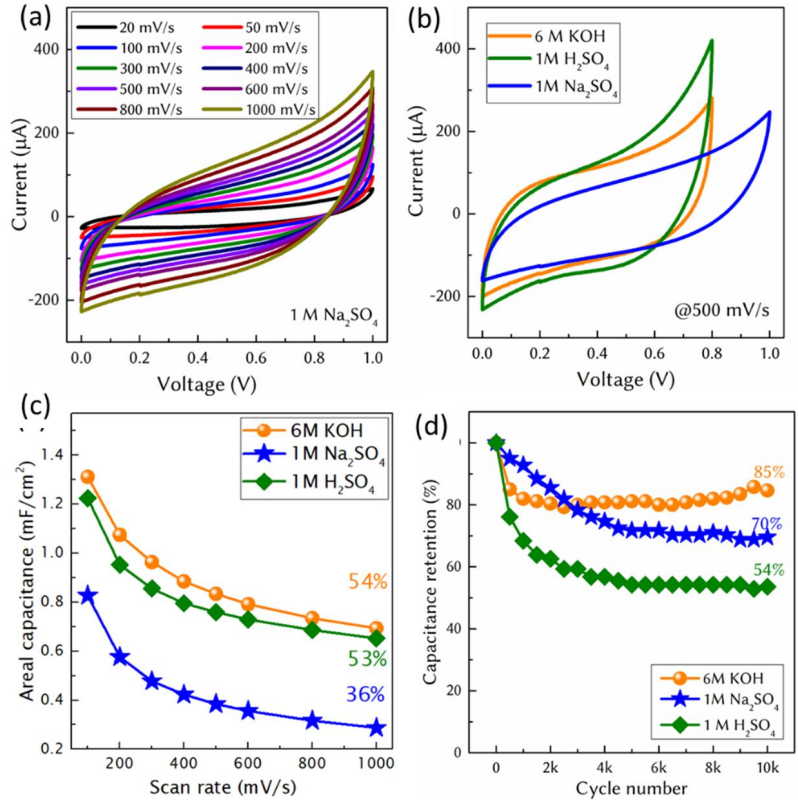

*Figure 2*: Charge-storage performance of carbyne. (a) Cyclic voltammogram of a symmetric device [3]Ph-cumulene in 1M Na$_2$SO$_4$, (b) comparative voltammogram of the devices in 1M Na$_2$SO$_4$, 1M H$_2$SO$_4$, and 6M KOH. Aqueous electrolyte dependent (c) areal capacitance with respect to the scan rate and their (d) cycle stability.

Evaluating the rate performance shows that the composite in 6M KOH, 1M H$_2$SO$_4$ and 1 M Na$_2$SO$_4$ maintain 54%, 53% and 36% at 1 V/s, respectively, compared to the areal capacitance measured at 0.1 V/s (36%). The charge-discharge profiles of the devices at different current densities are shown in Figure S1d-f, and the profiles are found to be symmetric and near-linear for all the cases, which is in good agreement with the CV results. Interestingly, the composite in 6M KOH stands out as more stable when compared to the other two after 10000 charge-discharge cycles. A closer inspection indicates an increased cycle stability of composite in 6M KOH over prolonged charge-discharge cycles after an initial decrease, which can be attributed to the chemical activation of electrode materials during the charge-discharge process.

The better rate performance and cycle stability of the composite in 6M KOH compared to that in 1M H$_2$SO$_4$ could be due to the balance between the contribution arises from double layer capacitance and pseudo-capacitance. Conducting polymer-based materials (PMMA in our case) are well known to contribute the pseudo-capacitance, and carbon-based materials (e.g., cumulenic *sp*-carbon here) provide double-layer capacitance. In order to estimate both contributions, using the partition procedure by Trasatti method analysis,[41][42] double layer capacitance is estimated from a plot of $C_A$ vs $v^{-1/2}$ (Figure 3a), and the total capacitance is estimated from the plot of $1/C_A$ vs $v^{1/2}$ (Figure 3b). The pseudo-capacitance is estimated from the difference between total capacitance and double layer capacitance. Interestingly, it has been seen that the composite provides higher double layer capacitance contribution in 6M KOH (Figure 3c) than that in 1M H$_2$SO$_4$, and, hence, better stability is obtained.

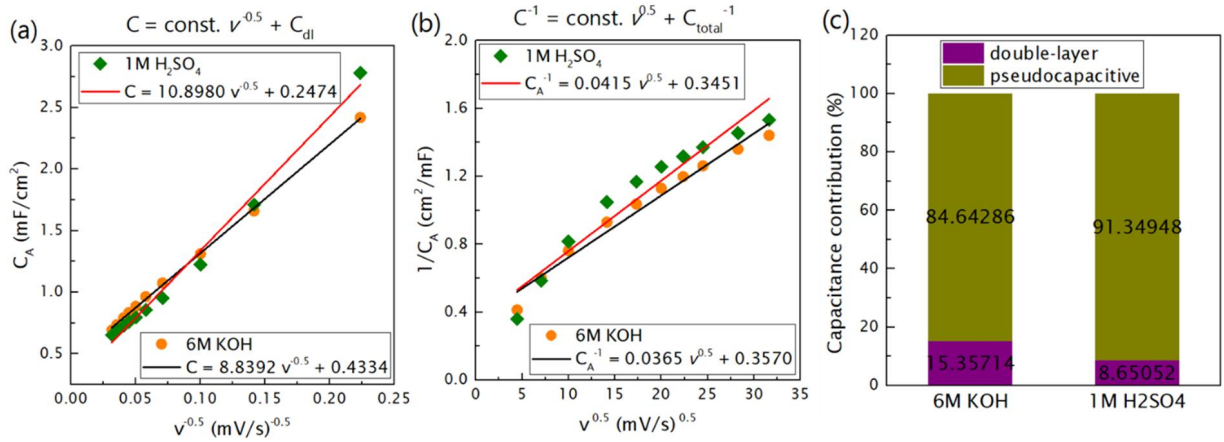

*Figure 3*: Plot of (a) areal capacitance ($C_A$) versus reciprocal square roots of scan rate ($v^{-1/2}$), and (b) reciprocal of areal capacitance vs $v^{1/2}$. (c) Estimated double layer capacitance and pseudocapacitance contribution of the composite in 6M KOH and 1M $H_2SO_4$. Solid lines in (a–b) represent linear fitting and the corresponding fitting equation is provided as an inset. $C_{dl}$ and $C_{total}$, presented in the equation (on the top of figure a–b) are the double-layer and total capacitance of the device, respectively.

Spectroscopic investigation of electrochemical impedance is carried out to probe further on the charge-storage kinetics, frequency response, and possibility of checking the utilization for applications for AC line filtering. The characteristics needed for filtering are a small resistor-capacitor time constant ($\tau_{RC}$), small characteristic relaxation time constant at –45° ($\tau_0$), large areal capacitance, and the small phase angle at 120 Hz (2nd harmonic of 60 Hz AC frequency, US standard). The characteristic relaxation time constant at –45° ($\tau_0$) represents the time needed when the discharge of the device goes 50% of maximum efficiency, where –45° is the boundary between capacitive and resistive behaviour.[43] The resistor-capacitor time constant ($\tau_{RC}$) is the time needed to charge 63.2% of the full potential of the capacitor.

From the Nyquist plot (Figure 4a), the lowest equivalent series resistance (ESR) of composite is achieved in 6M KOH (0.8 Ω) compared to that in 1M $Na_2SO_4$ (4.6 Ω), and 1M $H_2SO_4$ (1.7 Ω). Moreover, no semicircle is observed in the high-frequency zone of Nyquist plot indicates negligible charge-transfer resistance and hence excellent Ohmic contact between the electrode materials and current collector, fast electrochemical reaction rate and/or fast charge transfer kinetics for all devices represent the.[38] The Bode plot for each device is shown in Figure 4b. At 120 Hz, the negative phase angle for the composite in 6M KOH, 1M $H_2SO_4$, and 1M $Na_2SO_4$ is 52.8°, 51°, and 50°, respectively. Although the negative phase angles of our device at 120 Hz is lower than the ideal capacitor (90°) or vertically oriented graphene (85°),[44] it is comparable and/or better than *N*-doped mesoporous carbon (0°),[45] activated carbon-based commercial electric double layer capacitor (0°),[46][47] thermally reduced graphene oxide (30°)[48], 3D ordered porous graphene film (53°)[49], and carbon nanowall foam (47°).[50] Moreover, it has been demonstrated for $NiTe_2$-based devices that the negative phase angle around 53° is sufficient for the AC line filtering purposes.[51] The areal capacitance at 120 Hz of composite ($C_{dl}$) obtained in 1M $Na_2SO_4$, 1M $H_2SO_4$, and 6M KOH are 22, 66, and 177 µF/cm², respectively (Figure 4c). It has been reported that by reducing the thickness of a carbon nanotube film from 300 nm (loading ~32 µg/cm²) to 50 nm (~4 µg/cm²), the negative phase angle increases from 70.6° to 82.7°, and the areal capacitance decreases from 233 to 43 µF/cm². These results also indicate that using a thinner separator provides higher areal capacitance and negative phase angles.[52] In our case, the mass loading of material is much higher (1 mg/cm²) than in reference [52]. Decreasing the electrode thickness may be the plausible solution to increase the negative phase angle at 120 Hz,[53] which is in the scope of our future research. The maximum areal capacitance obtained for our composite in 1M $Na_2SO_4$, 1M $H_2SO_4$, and 6M KOH are 0.42, 0.79 and 2.63 mF/cm$^2$, respectively at the lowest frequency (Figure 4c).

Time constants $\tau_0$ and $\tau_{RC}$ are estimated for all devices from the impedance spectroscopic data. The estimated $\tau_0$ for KOH, $H_2SO_4$, and $Na_2SO_4$ are 0.7, 1.3 and 1 ms, respectively and the corresponding characteristics frequency ($f_0$) at −45° are 1301, 756, and 1000 Hz. The capacitor-resistor time constant of our devices is 1.14, 1.04, and 1.04 ms in KOH, $H_2SO_4$, and $Na_2SO_4$, respectively, which is ~3 orders of magnitude smaller than that of the commercial ECs (~1 second).[54] The small values of $\tau_{RC}$ and $\tau_0$ for our devices suggests ultrafast frequency response.

The real part of impedance (Z') derives of ion migration, electrical conduction, and ion diffusion. The slope of the curve represents the diffusion impedance. Since the active material here is the same, [3]Ph-cumulene-PMMA composite, electrical conductivity is assumed to be same for all identical electrodes prepared by the same protocol. Thus, the differences in charge-storage performances and kinetics in three different electrolytes are attributed, mainly, to the electrolyte ion characteristics. The ion migration in supercapacitor takes place at low frequency region (below 100 Hz), and the diffusion occurs in the mid-frequency region.[43] Here, we estimated the ionic diffusion coefficient from the plot of Z' vs $\omega^{-0.5}$ (Figure 4d). The slope of the curve is called as the Warburg coefficient (σ).[55] The estimated σ-values of electrode in 6M KOH, 1M $H_2SO_4$, and 1M $Na_2SO_4$ are 198, 554, and 1720 Ω/s$^{0.5}$, respectively. This suggests that 6M KOH electrolyte is more compatible with the composite to have better ion diffusion.

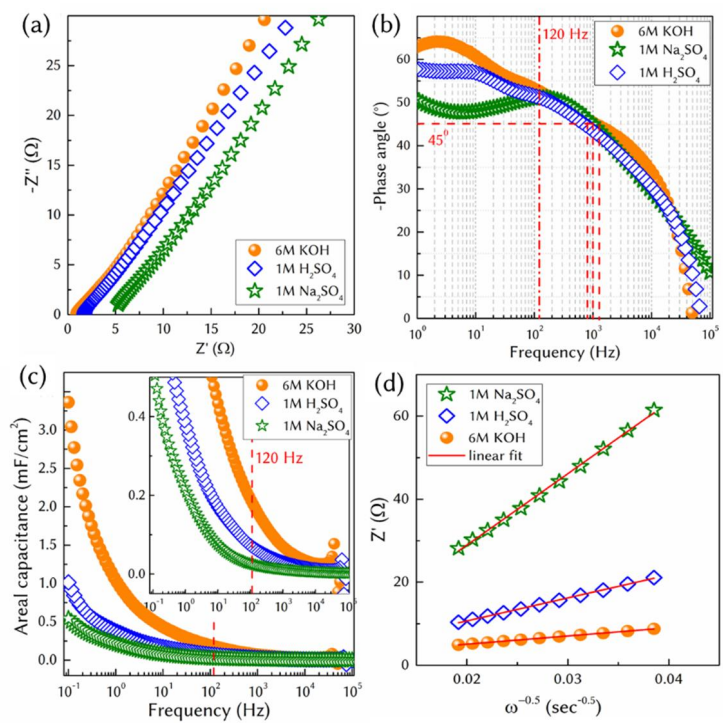

Figure 4: Frequency response. (a) Nyquist plot, (b) Bode plot, (c) frequency dependent areal capacitance, and (d) plot of Z' vs $\omega^{-0.5}$ of [3]Ph-cumulene supercapacitor in 1M $Na_2SO_4$, 1M $H_2SO_4$, and 6M KOH.

The stability of the [3]Ph-cumulene-based supercapacitors in the aggressive electrolytes has been verified by comparing the Raman spectra of the electrodes before and after electrochemical measurements (Figure 5a). The spectra are all similar, and the Raman peaks of the [3]Ph-cumulene remain effectively unchanged after immersion in the different electrolytes. These results confirm that the PMMA matrix is highly effective in the stabilization and protection of the [3]Ph-cumulene even in highly alkaline and basic environments. A more detailed investigation has been done considering the relative area of the ECC peak of [3]Ph-cumulene (2035 cm$^{-1}$) with respect to the area of the signal of the phenyl group (1593 cm$^{-1}$) in thin films before and after the electrochemical measurements. Variations in this value would document degradation of [3]Ph-cumulene during the experiment. Figure 5b shows that the relative ratio of signal ECC/phenyl is found to be between 2.12 to 2.3 for the [3]Ph-cumulene before and after the electrochemical investigations. The relative reduction in the ECC signal is only 4.3% and 7.2% in the case of 1M $H_2SO_4$ and 6M KOH, respectively, and emphasizes the excellent structural stability of *sp*-carbon as electrodes.

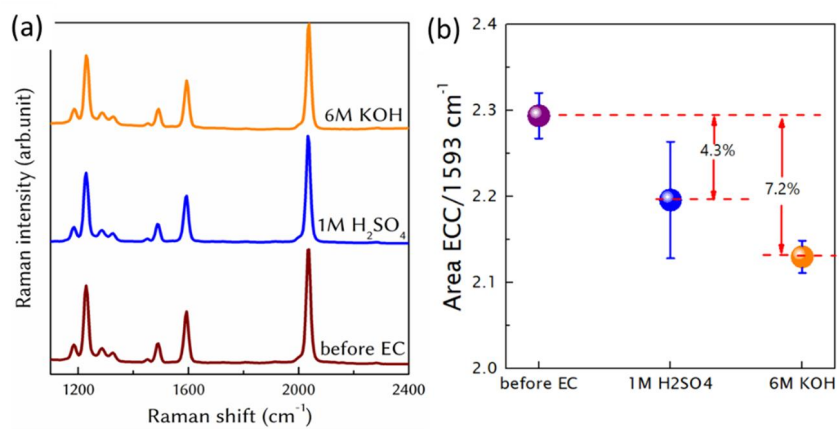

*Figure 5: (a) Raman spectra and (b) evolution of the ratio of Raman signals of ECC/phenyl (at 199x and 1593 cm$^{-1}$, respectively) of [3]Ph-cumulene-PMMA composites before and after the electrochemical measurements in the different aqueous electrolytes.*

CONCLUSION

We have investigated the electrochemical charge-storage properties and frequency response of a PMMA-wrapped cumulenic *sp*-carbon atomic wires in supercapacitor devices using three different aqueous electrolytes, namely, 1M $Na_2SO_4$, 1M $H_2SO_4$, and 6M KOH. The electrode material delivered higher areal capacitance with fast frequency response, rate performance, cycle stability, higher negative phase angle, and smaller time constants in 6M KOH compared to that in 1M $Na_2SO_4$ and 1M $H_2SO_4$. This performance is attributed to the better ionic diffusion, and balanced double-layer capacitance and pseudo-capacitance in 6M KOH electrolyte. Analysis after electrochemical measurements confirmed that the active component, [3]Ph-cumulene, does not appreciably degrade, even under the rigorous conditions of electrochemical charge/discharge. Thus, [3]Ph-cumulenic *sp*-carbon atomic wires hold great promise as an electrode material for electrochemical energy storage and AC-line filtering applications.

EXPERIMETAL DETAILS

The [3]Ph-cumulene electrode is synthesized using a blend of PMMA/[3]Ph-cumulene in N,N-Dimethylformamide (DMF, HPLC grade ≥99.7%, Alfa Aesar). The composition of these films was optimized to achieve the highest possible concentration of [3]Ph-cumulene relative to PMMA while maintaining a minimum amount of polymer. The mixture was prepared by dissolving 0.02 g of PMMA powder in 1 mL of DMF, achieving complete dissolution through stirring at 50 °C for three hours. Separately, a second solution containing the cumulene was prepared by dissolving 8 mg of [3]Ph-cumulene powder in 1 mL of DMF, ensuring homogeneity through stirring and heating at 40 °C for three hours. To maximize the concentration of [3]Ph, this solution was brought close to its solubility limit in DMF. The two homogeneous solutions were then mixed and stirred for three hours at 30 °C. A drop of the resulting solution was drop-cast onto a carbon-paper substrate, and gentle evaporation of the solvent was achieved by placing the sample under a vented chemical hood. After complete solvent evaporation, the drop-casting process was repeated until a film of sufficient thickness (> 1 mg/cm$^2$) was obtained.

Raman spectra of the pure powder of [3]Ph-cumulene and of [3]Ph-cumulene embedded within the PMMA film – both before and after electrochemistry measurements - have been acquired with a Renishaw inVia Raman microscope with a diode-pumped solid-state laser with two possible wavelengths (λ=532 or 660 nm). All the spectra were acquired with a 532 nm excitation laser and with 500 accumulations of 0.1 s each. The laser power was set to 0.7 mW to avoid degradation of the cumulene during the measurements. To address the inhomogeneities in the cumulene-PMMA nanocomposites, spectra were averaged by measuring at various spots across the film.

UV-Vis absorption spectra of [3]Ph-cumulene were recorded dissolving the cumulene powder in DMF. The final concentration was about 10$^{-5}$ M. Spectra were recorded at room temperature using a Shimadzu UV-1800 UV/Visible scanning spectrophotometer with a detection range of 190-1100 nm. The sampling interval of the spectra was set to 0.2 nm. The sample solution was analyzed inside a quartz cuvette with a 10 mm optical path.

The electrochemical charge-storage performances of the samples were conducted in Swagelok Cell (SKU: ANR-B01, Singapore) and three standard aqueous electrolytes, namely 1M Na$_2$SO$_4$, 1M H$_2$SO$_4$, and 6M KOH, are used. The hydrophobic PP membrane is modified by a two-step process: soaked with acetone at 20 °C for 5 min and followed by aqueous 6M KOH solution at 20 C, and used after 1 hr.[51] The cell was assembled by sandwiching separator-soaked-electrolytes between as-prepared electrode material grown on carbon paper. For the electrochemical testing, electrode materials and modified separator were dipped into the electrolyte solution for 1hr. Cyclic voltammogram, charge-discharge test, and electrochemical impedance spectra were recorded using a PALMSENS electrochemical workstation. The cyclic voltammetry at different scan rates ranging from 20 to 1000 mV/s and charge-discharge at different current densities of 150 to 500 µA were carried out and scanned within the defined voltage at 100 mV/s for 1000 times. The areal capacitance is calculated using the equation: $C_{areal} = \int I\, dV / A.v.\Delta V$, where $I$ is the current, $v$ is the scan rate, $A$ is the geometric area of the electrode and $\Delta V$ is the voltage of the device. Single electrode capacitance = 4 × device capacitance. The volumetric capacitance of electrode materials is estimated by dividing the areal capacitance by the total height of two carbon nanofoam electrodes. The electrochemical impedance spectroscopy is conducted in the frequency range of 1 Hz to 0.1 MHz at open circuit potential with a 10 mV *a.c.* perturbation. The energy density ($E_A$) of the device is calculated via the equation: $E_A = C_{dl}V^2/2$, where $C_{dl}$ is the double layer capacitance at 120 Hz obtained from the areal capacitance versus frequency plot. The relaxation time constant is calculated from the impedance spectra at 120 Hz using the equation: $\tau_{RC} = -Z'/2\pi f Z''$, where Z' and Z'' are the real and imaginary components of impedance.

**Supporting information.**

Optical photograph of the design of the set-up. Additional SEM, XPS, Raman spectra and fitted curves. Additional electrochemical testing results.

**Authors contributions**

S.G. and C.S.C planned and conceptualized the work. S. M. did the electrode material preparation, UV-Vis spectroscopy and Raman spectroscopy measurement, and assisted in manuscript preparation. S. G. and M. R. did the electrochemical measurements. R.R.T. and Y.Q. conducted the syntheses and purification of [3]Ph-cumulene. S. G. wrote the draft. All authors edited the manuscript and approved the final version of the manuscript.

**Notes**

The authors declare no competing financial interest.

ACKNOWLEDGEMENT

S.G thanks Horizon Europe (HORIZON) for the Marie Sklodowska-Curie Fellowship (grant no. 101067998-ENHANCER). C.S.C. acknowledges partial funding from the European Research Council (ERC) under the European Union's Horizon 2020 Research and Innovation Program ERC Consolidator Grant (ERC CoG2016 EspLORE Grant Agreement 724610, website: www.esplore.polimi.it). C.S.C. also acknowledges funding by the project funded under the National Recovery and Resilience Plan (NRRP), Mission 4 Component 2 Investment 1.3 Call for Tender 1561 of 11.10.2022 of Ministero dell'Università e della Ricerca (MUR), funded by the European Union NextGenerationEU Award Project Code PE0000021, Concession Decree 1561 of 11.10.2022 adopted by Ministero dell'Università e della Ricerca (MUR), CUP D43C22003090001, Project "Network 4 Energy Sustainable Transition (NEST)". R.R.T. acknowledges funding from the Natural Sciences and Engineering Research Council of Canada (RGPIN-2023-04000) and the Canada Foundation for Innovation (CFI)

DATA AVAILABILITY: All the data of this study are available in the main manuscript and the Supplementary Information.

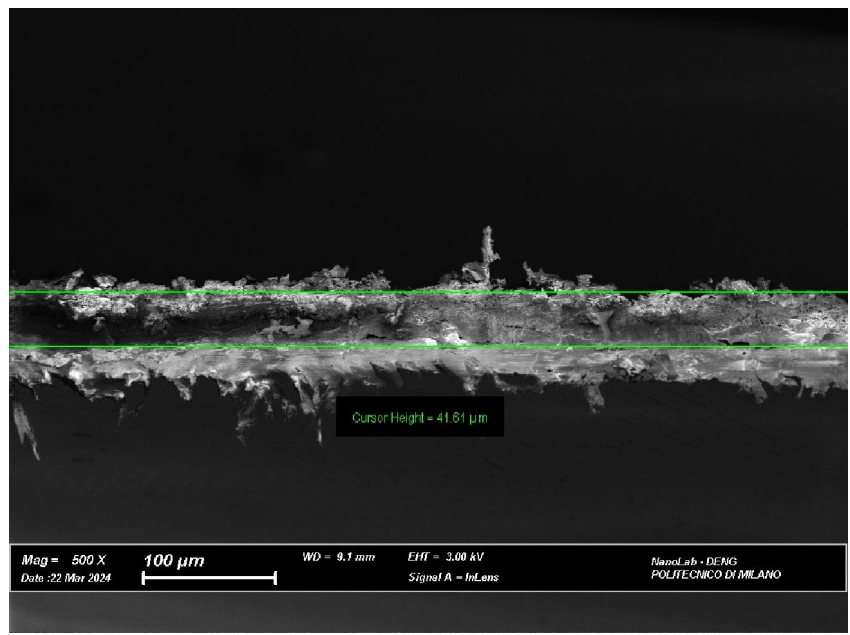

*Figure S2:* Cross-sectional scanning electron micrograph of [3]ph-cumulenic sp-carbon atomic wires/PMMA composite

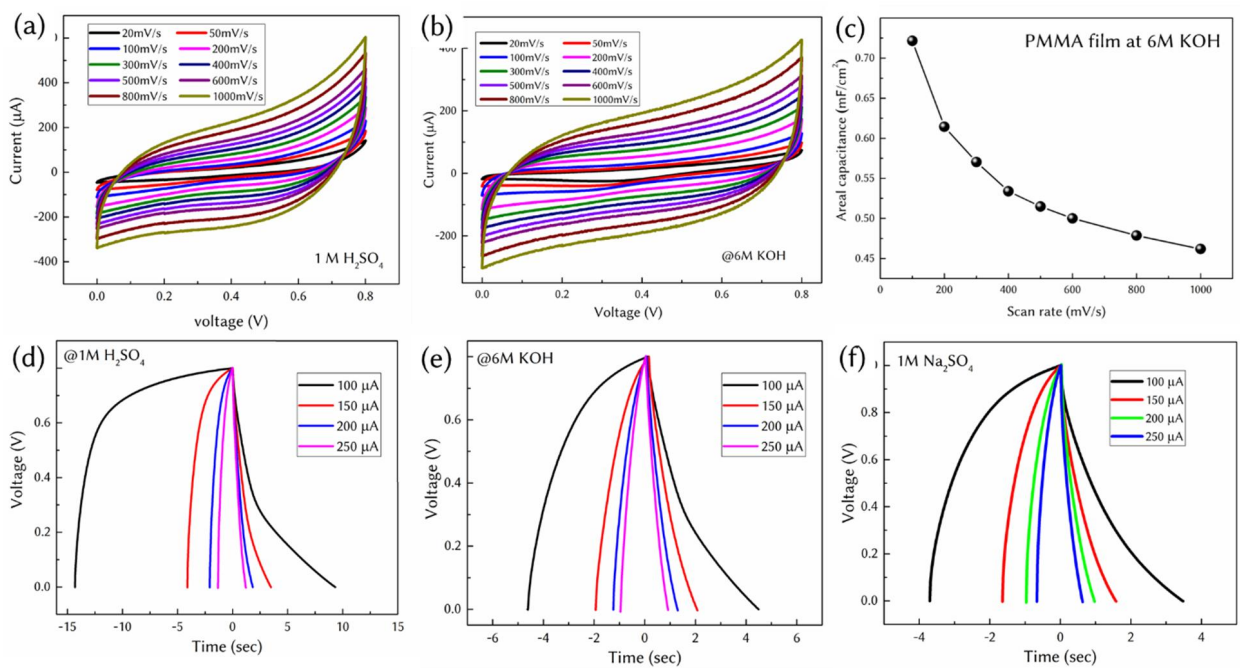

*Figure S2: Cyclic voltammogram of composite in (a) 1M $H_2SO_4$ and (b) 6M KOH. Areal capacitance of PMMA film-based symmetric device with respect to the scan rate in 6M KOH electrolyte. Charge-discharge profile at different current densities for the composite in (d) 1M $H_2SO_4$, (e) 6M KOH and (f) 1M $Na_2SO_4$*

17